\documentclass[]{revtex4-2}

\usepackage{graphicx}
\usepackage{dcolumn}
\usepackage{bm}
\usepackage{amsmath}
\usepackage{caption}
\usepackage{subcaption}
\usepackage{enumitem}
\usepackage{tikz}
\DeclareRobustCommand{\bigcnum}[1]{%
  \tikz[baseline=(c.base)]{
    \node[circle,draw,inner sep=1.1pt,line width=0.2pt] (c) {\scriptsize #1};
  }
}
\usepackage{textcomp}

\begin{document}

\preprint{APS/123-QED}

\title{Reconstructing a large-scale matter-density contrast profile to reconcile Pantheon+ supernovae with DESI DR2 BAO in an inhomogeneous universe}

\author{Toshifumi Futamase}
\email{tof@astr.tohoku.ac.jp}
\affiliation{Astronomical Institute, Tohoku University, Sendai, Miyagi,980-8578, Japan}%
\altaffiliation[Also at ]{Academia Sinica Institute of Astronomy and Astrophysics (ASIAA), No.~1, Section~4, Roosevelt Road, Taipei 10617, Taiwan}

\author{Reiki Kojima}%
 \email{kojima\_r@u.phys.nagoya-u.ac.jp}
\author{Masanori Tomonaga}
\email{tomonaga.masanori.b3@s.mail.nagoya-u.ac.jp}
\affiliation{Graduate School of Science, Nagoya University, Furo-cho, Chikusa-ku, Nagoya, Aichi 466-8602, Japan}

\date{\today}

\begin{abstract}
    The Hubble parameters measured by the DESI DR2 BAO observations show a significant discrepancy from the prediction of the standard cosmological model.
    This discrepancy, together with the long-discussed Hubble tension, may originate from large-scale inhomogeneities in the matter distribution.
    This interpretation is motivated by infrared galaxy surveys, which suggest that our galaxy resides within the $\sim300$ Mpc under-dense region known as the KBC void.
    In this study, we apply a linear order relation -- relating the horizon-scale Hubble parameter inferred from CMB observations and the local-scale Hubble parameter -- to the Pantheon+ Type Ia supernovae and the DESI DR2 BAO data.
    We show that a simple inhomogeneous cosmological model consisting of eight top-hat shells can consistently explain the Hubble parameters inferred from both observations.
    Based on the matter-density distribution, we also briefly discuss its possible impact on cosmological observables, including the magnitude--redshift relation, the kinematic Sunyaev--Zel'dovich effect, and the integrated Sachs--Wolfe effect.
\end{abstract}

\maketitle

\section{\label{sec:level1}Introduction}
    The Pantheon+ combined with the SH0ES Cepheid host distance calibration reports a Hubble constant of \(H_0 = 73.6 \pm1.1\,\mathrm{km\,s^{-1}\,Mpc^{-1}}\) for a flat $\Lambda$CDM model \cite{brout2022,riess2022}.
    In contrast, the Planck 2018 results based on CMB observations indicate 
    $
    H_0 = 67.4 \pm 0.5\ \mathrm{km\,s^{-1}\,Mpc^{-1}}
    $
    \cite{planckcollaboration2020}. 
    The discrepancy between these two values corresponds to a significant tension $\sim 5\sigma$, commonly referred to as the Hubble tension (see \cite{verde2019a,verde2024} for a review).
    
    Two main approaches have been proposed to alleviate this tension. 
    One is to modify the physics of the early universe, such as in Early Dark Energy (EDE) models, which reduce the sound horizon and lead to a higher inferred expansion rate (reviewed in \cite{poulin2023}). 
    The other is the local void scenario, in which a local underdensity induces an effective curvature \cite[e.g.,][]{enqvist2008}, enhancing the locally measured expansion rate. 
    This hypothesis is supported by several observations, including the $\sim300\ \mathrm{Mpc}$ underdensity with contrast $\delta\simeq -0.3$ (the Keenan-Barger-Cowie (KBC) void; \cite{keenan2013}), a $\sim200\ \mathrm{Mpc}$ region with $\delta\simeq -0.2$ \cite{wong2022}, and a $\sim100\ \mathrm{Mpc}$ region with $\delta\simeq -0.3$ \cite{bohringer2020}.
    
    While these discussions have predominantly focused on the local universe at $z \lesssim 1$, the Dark Energy Spectroscopic Instrument (DESI) has recently provided measurements of the cosmic expansion history over a wider redshift range, $0.29 \lesssim z \lesssim 2.33$, using the baryon acoustic oscillation (BAO) distances in galaxies, quasars and Lyman-$\alpha$ forest \cite{desicollaboration2025}. 
    However, the Hubble parameters inferred from both the radial BAO (Hubble distance, $D_\mathrm{H}$) and the transverse BAO (comoving angular diameter distance, $D_\mathrm{M}$) measurements disfavor the predictions of the $\Lambda$CDM model \cite{efstathiou1990,ostriker1995} in several redshift ranges.
    
    We interpret this observed discrepancy, namely that the expansion rates inferred from Pantheon+ (SNe Ia) \cite{brout2022} and DESI DR2 (BAO) \cite{desicollaboration2025} are inconsistent with the CMB-based constraints from Planck 2018, within the framework of local inhomogeneities.
    In such an inhomogeneous universe, observables are, in principle, appropriately treated using light-cone averaging \cite{nugier2013}.
    This is because, although the universe is intrinsically highly inhomogeneous, it can be effectively described as homogeneous and isotropic as a result of an averaging procedure along the past light cone.
    However, due to the complexity in the correspondence between the light-cone average and cosmological parameters such as the Hubble parameter, we adopt the spatial averaging procedure on a constant-time hypersurface \cite{buchert2000,buchert2001a,buchert2020}.
    Refs.~\cite{kasai2019,tomonaga2021,tomonaga2023} provide an equation relating the horizon-scale Hubble parameter inferred from CMB observations and the local-scale Hubble parameter within the framework of spatial averaging.
    The main objective of this paper is to roughly identify a spatially averaged mass-density contrast profile that is simultaneously consistent with the expansion rate inferred from Pantheon+ Type Ia supernovae (SNe Ia) and that obtained from the DESI DR2 BAO data over a wide redshift range of $0 \lesssim z \lesssim 2.33$ based on the relation.
    
    The structure of this paper is as follows.
    In Section~\ref{sec:averaging}, we briefly review the linear-order relation that connects the horizon-scale Hubble parameter inferred from CMB observations with the local-scale Hubble parameter within the framework of spatial averaging.
    In Section~\ref{sec:method}, we describe the method for reconstructing the large-scale matter-density contrast distribution based on observations.
    In Section~\ref{sec:diccussion}, we present the reconstructed large-scale matter-density contrast distribution and provide a brief discussion of its impact on the magnitude--redshift ($m$--$z$) relation, the kinematic Sunyaev--Zel'dovich (kSZ) effect, and the integrated Sachs--Wolfe (ISW) effect.
    
    We use the following convention: Greek indices $\mu,\nu,\ldots$ run from $0$ to $3$, and Latin indices $i,j,k,\ldots$ run from $1$ to $3$.

\section{Volume expansion over a finite domain in an inhomogeneous universe}\label{sec:averaging}
    In this section, we briefly summarize the framework of cosmology in which the volume expansion is derived by the spatially averaged matter density.
    
    We consider the model which contains a pressure-less perfect fluid (dust), with the energy-momentum tensor given by:
    \begin{align}
        T^{\mu\nu} = \varrho\, u^\mu u^\nu,
    \end{align}
    with the matter density
    \begin{align}
        \varrho = \rho_b(t) \left( 1 + \delta(t, \bm{x}) \right).
    \end{align}
    \(u^\mu\) is the four-velocity of a comoving observer, and \(\delta(t, \bm{x})\) represents the matter-density contrast with respect to the background matter density \(\rho_b(t)\).

    We introduce the spatial average of a scalar quantity \(Q\) over a domain \(\mathcal{D}\) defined as
    \begin{align}
        \langle Q \rangle := \frac{1}{V_\mathcal{D}} 
        \int_{\mathcal{D}} Q \sqrt{\det(\gamma_{ij})} \, d^3x \label{eq:spatial_average},
    \end{align}
    where \(V_\mathcal{D}\) is the spatial volume of the domain \(\mathcal{D}\) on a constant-time hypersurface \(\Sigma_t\),
    \begin{align}
        V_\mathcal{D} := \int_{\mathcal{D}} \sqrt{\det(\gamma_{ij})} \, d^3x .
    \end{align}
    Here, \(\gamma_{ij}\) denotes the induced spatial metric on \(\Sigma_t\).
    By applying the spatial averaging defined in Eq.~\eqref{eq:spatial_average} to the Einstein equations
    Eqs.~\eqref{eq:Einstein00}, \eqref{eq:Einstein0i}, and \eqref{eq:Einsteinij}
    over a finite domain $\mathcal{D}$, the domain can be regarded as an effective Friedmann universe
    (Here, a homogeneous and isotropic universe is referred to as a Friedmann universe).
    Within the framework of linear perturbation theory, the scale factor of the effective Friedmann universe
    is then given, as a scale factor depending on the spatially averaged domain $\mathcal{D}$, by
    \begin{equation}
    a_{\mathcal{D}}(z)
    = a(z)\left(1 - \frac{1}{3}\langle \delta \rangle(z)\right).
    \end{equation}
    In this case, the effective Friedmann equation is written as
    \begin{align}
    H^2_{\mathcal{D}}
    = \frac{8\pi G}{3c^2} \frac{\langle \varrho \rangle}{a_\mathcal{D}^3}
    - \frac{c^2 K_\mathrm{eff}}{a_\mathcal{D}^2}
    + \frac{c^2 \Lambda}{3},\label{eq:average_Friedman}
    \end{align}
    where $H_{\mathcal{D}} := \dot{a}_\mathcal{D}/a_\mathcal{D}$ is the Hubble parameter in the domain
    $\mathcal{D}$, and $K_\mathrm{eff}$ denotes the effective curvature of the domain.
    It is given by
    \begin{align}
    K_\mathrm{eff}
    = \frac{2}{3c^2}\frac{\langle\delta\rangle(z)}{D_+(z)}.
    \label{eq:Keff}
    \end{align}
    We note that this effective Friedmann universe can also be obtained in the same form
    by spatially averaging the LTB solution (\cite{lemaitre1997,tolman1997,bondi1947}) over the domain $\mathcal{D}$,
    as shown in Appendix~\ref{sec:LTB}.
    Furthermore, the spatially averaged expansion rate over the domain $\mathcal{D}$ is given by
    \begin{align}
    H_{\mathcal{D}}(z)
    = H(z)\left(1 - \frac{1}{3} f(z)\,\langle\delta\rangle(z) \right),
    \label{eq:local_global_hubble}
    \end{align}
    as shown in Refs.~\cite{kasai2019,tomonaga2021,tomonaga2023}.
    
    Here, \(\langle\delta\rangle(z)\) is the matter-density contrast spatially averaged over the domain \(\mathcal{D}\),
    \(f := d\ln D_+/d\ln a\) is the linear growth rate, where \(D_+(z) = H\int^t dt' /(aH)^2\) denotes the linear growth factor,
    and \(H(z)\) denotes the background (horizon-scale) Hubble parameter inferred from Planck~2018 within the standard cosmological model.
    Eq.~\eqref{eq:local_global_hubble} provides a linear-order relation between the horizon-scale Hubble parameter and the Hubble parameter spatially averaged over the domain \(\mathcal{D}\).
    Consequently, an under-dense (over-dense) region leads to an enhanced (suppressed) cosmic expansion relative to the Planck~2018 prediction.

\section{Reconstructing a Spatially Averaged Matter-Density contrast Distribution}\label{sec:method}
    In this section, we derive the spatially averaged matter-density contrast, $\left<\delta\right>(z)$, based on Eq.~\eqref{eq:local_global_hubble}, using the Hubble parameters inferred from the Pantheon+ SNe Ia and DESI DR2 BAO observations.
    To reconstruct the spatially averaged matter-density contrast, we assume a simple model in which large-scale matter inhomogeneities are represented by eight top-hat structures. 
    The redshift range of each top-hat shell is defined by centering it on the effective redshift of the observational data point, with the shell boundaries given by the midpoints between the effective redshifts of adjacent data points.
    Each structure is characterized by a single parameter representing its density contrast, resulting in a total of eight free parameters, which are determined through a $\chi^2$ minimization.
    
    We perform the $\chi^2$ fitting using a dataset of eight points, including seven independent observational data points (the local Hubble parameter $H_{\mathcal{D}}(z=0)$ from SNe and six Hubble parameters derived from $D_\mathrm{H}$ measurements of BAO) and one additional pseudo–observational data point derived from the model constrained by these seven independent points, together with the isotropic BAO distance $D_\mathrm{V}$. The procedure for deriving $\left<\delta\right>(z)$ is summarized as follows.

\begin{enumerate}[label=(\roman*), leftmargin=*, align=left] 
  \item Computation of the Hubble parameters at $z_\mathrm{eff} =0, 0.510, 0.706, 0.934, 1.321, 1.484$ and $2.330$ from SNe and radial BAO distance $D_\mathrm{H}$
    \begin{itemize}[leftmargin=*, label={--}]
      \item As the local expansion rate at the effective redshift \(z_\mathrm{eff}=0\),  
      we adopt \(H_0 = 73.6 \pm 1.1\,\mathrm{km\,s^{-1}\,Mpc^{-1}}\), as determined by Pantheon+ based on SH0ES SNe Ia observations \cite{brout2022}.
      \item For the effective redshifts \(z_\mathrm{eff} = 0.510, 0.706, 0.934, 1.321, 1.484\) and \(2.330\), we derive the Hubble parameters from the radial BAO distance \(D_\mathrm{H}\) listed in Table~\ref{tab:DESI} using the relation
        \begin{align}
            D_\mathrm{H}(z) &= \frac{c}{H_{\mathcal{D}}(z)}.\label{eq:DESI_DH}
        \end{align}
    \end{itemize}

  \item Estimation of the Hubble parameter at $z_\mathrm{eff} =0.295$ using the isotropic BAO distance $D_\mathrm{V}$
    \begin{itemize}[leftmargin=*, label={--}]
        \item Using the seven Hubble parameters obtained in step~(i), we obtain an initial estimate of the matter-density contrast profile $\langle \delta\rangle(z)$ through Eq.~\eqref{eq:local_global_hubble}.
    
        \item Based on this initial matter-density contrast profile, we interpolate the $K_\mathrm{eff}$ and $H_\mathcal{D}$ as a function of redshift using Eqs.~\eqref{eq:Keff} and \eqref{eq:local_global_hubble}.
        Using these quantities, we compute the transverse comoving distance at $z_\mathrm{eff}=0.295$ as
        \begin{align}
            D_\mathrm{M} &=
            \begin{cases}
                \dfrac{1}{\sqrt{-K_{\mathrm{eff}}}}
                \sinh\!\left(c\,\sqrt{-K_{\mathrm{eff}}}
                \displaystyle\int_0^z \frac{dz^\prime}{H_{\mathcal{D}}(z^\prime)}\right),
                & K_{\mathrm{eff}}<0, \\[12pt]
                \dfrac{1}{\sqrt{K_{\mathrm{eff}}}}
                \sin\!\left(c\,\sqrt{K_{\mathrm{eff}}}
                \displaystyle\int_0^z \frac{dz^\prime}{H_{\mathcal{D}}(z^\prime)}\right),
                & K_{\mathrm{eff}}>0, \\[12pt]
                \displaystyle\int_0^z \frac{c\,dz^\prime}{H_{\mathcal{D}}(z^\prime)},
                & K_{\mathrm{eff}}=0.
            \end{cases}
        \end{align}

        \item
        Using the transverse comoving distance $D_\mathrm{M}$ and the observed
        isotropic BAO distance $D_\mathrm{V}$ at $z_\mathrm{eff}=0.295$
        listed in Table~\ref{tab:DESI}, we estimate $D_\mathrm{H}$ and derive
        the corresponding pseudo–observational value of
        $H_{\mathcal D}(z)$ using Eq.~\eqref{eq:DESI_DH}, where
        \begin{align}
        D_\mathrm{V} := (z D_\mathrm{M}^2 D_\mathrm{H})^{1/3}.
        \end{align}
    \end{itemize}

  \item Determination of the matter density contrast-parameters of each shell
    \begin{itemize}[leftmargin=*, label={--}]
      \item We determine the eight parameters of matter-density contrast by minimizing
      \begin{align}
        \chi^2 = \sum_{i=1}^{8}\frac{\left(H^{\mathrm{obs}}_i - H^{\mathrm{model}}_i\right)^2}{\left(\sigma_i^{\mathrm{obs}}\right)^2},
      \end{align}
      where $H_i^{\mathrm{model}}$ denotes the Hubble parameters computed from Eq.~\eqref{eq:local_global_hubble} using the reconstructed matter density contrast profile,
       and $H_i^{\mathrm{obs}}$ denotes the Hubble parameters obtained from the procedures (i) and (ii).
    \item The $1\sigma$ uncertainties of the density parameters obtained from the $\chi^2$ fitting are estimated using 1000 Monte Carlo realizations.
    \end{itemize}

\end{enumerate}

    \begin{table}
    \centering
    \caption{
    Constraints on the BAO distance ratios at the effective redshifts $z_{\mathrm{eff}}$,
    taken from Table~IV of \cite{desicollaboration2025}.
    Here, $r_\mathrm{d}$ denotes the sound horizon at the drag epoch.
    }
    \begin{tabular}{c c c}
    \hline\hline
    $z_{\mathrm{eff}}$ & $D_\mathrm{V}/r_\mathrm{d}$ & $D_\mathrm{H}/r_\mathrm{d}$ \\
    \hline
    0.295 & $7.942 \pm 0.075$ & -- \\
    0.510 & -- & $21.863 \pm 0.425$ \\
    0.706 & -- & $19.455 \pm 0.330$ \\
    0.934 & -- & $17.641 \pm 0.193$ \\
    1.321 & -- & $14.176 \pm 0.221$ \\
    1.484 & -- & $12.817 \pm 0.516$ \\
    2.330 & -- & $8.632 \pm 0.101$ \\
    \hline
    \end{tabular}
    \label{tab:DESI}
    \end{table}

    \section{Results and Discussion}\label{sec:diccussion}
\begin{figure*}
        \centering
        \begin{subfigure}[t]{0.48\textwidth} 
            \centering
            \includegraphics[width=\linewidth]{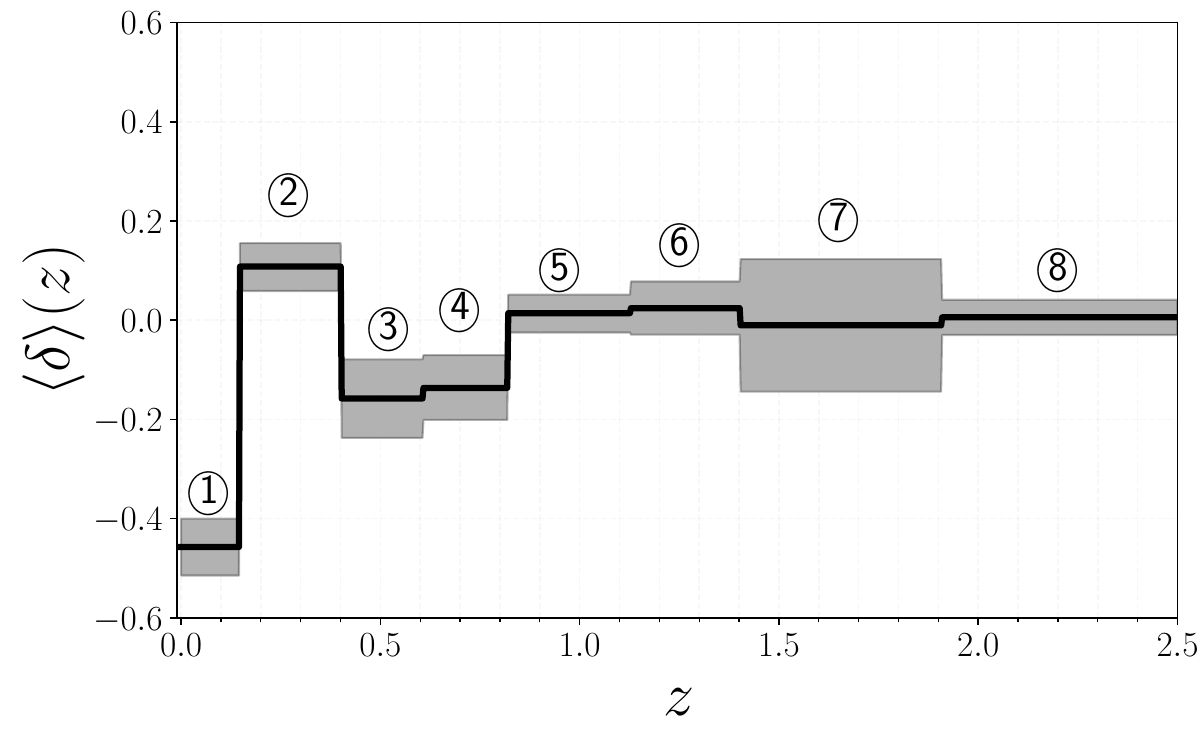}
            \subcaption{Spatially averaged mass-density distribution, consisting of eight top-hat structures (See Table~\ref{tab:voids}).}
        \end{subfigure}
        \hfill
        \begin{subfigure}[t]{0.48\textwidth} 
            \centering
            \includegraphics[width=\linewidth]{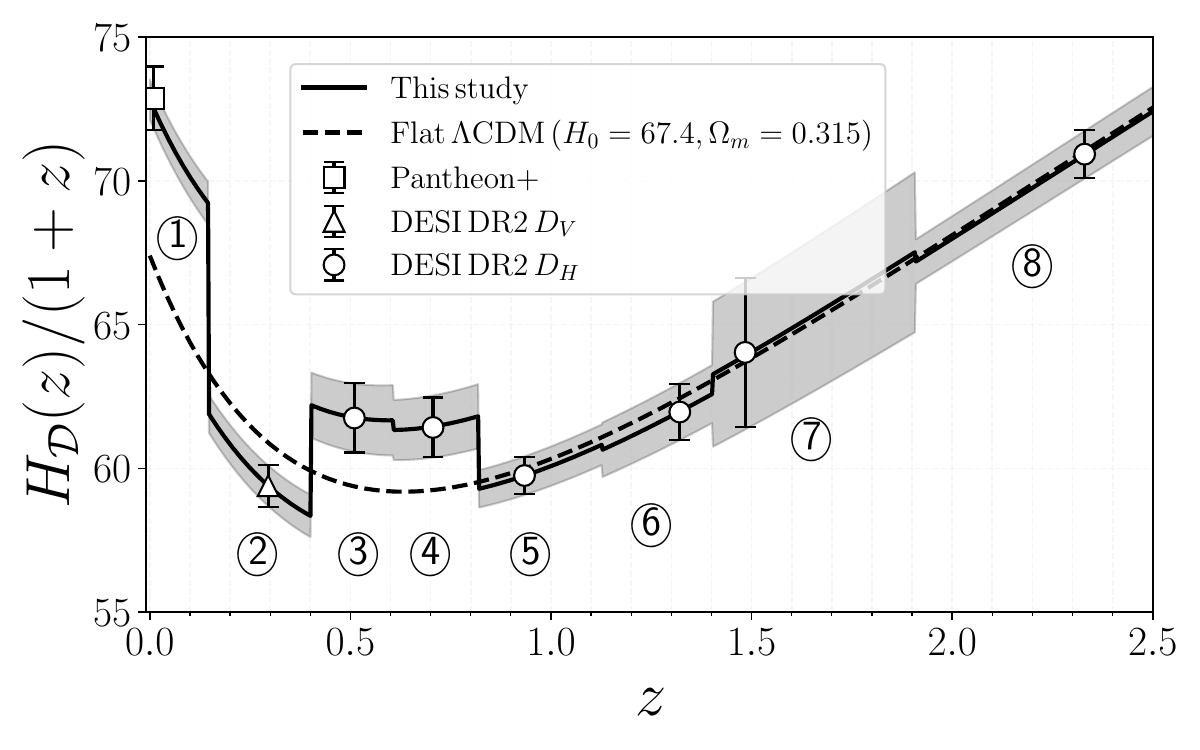}
            \subcaption{The redshift dependence of $\dot{a}_{\mathcal{D}}= H_{\mathcal{D}}(z)/(1 + z)$ inferred from the Pantheon+ SNe Ia and the radial and isotropic BAO distances $D_\mathrm{H}$ and $D_\mathrm{V}$.}
        \end{subfigure}
    
        \caption{
            Reconstruction of the matter density distribution and its impact on the expansion history.
            In both panels, the horizontal axis represents the redshift range over which the spatial averaging is performed.
            The gray shaded regions indicate the 68\% (1$\sigma$) credible intervals.
        }
        \label{fig:fourfigs}
    \end{figure*}

    Fig.~\ref{fig:fourfigs}(a) shows the reconstructed distribution of the spatially averaged matter-density contrast over the redshift range $0 \lesssim z \lesssim 2.33$. 
    Table~\ref{tab:voids} summarizes the spatially average matter-density contrast in each region that constitutes the reconstructed matter-density contrast distribution shown in Fig.~\ref{fig:fourfigs}(a).
    The innermost local region (\bigcnum{1}) corresponds to a KBC-void-like under-dense region. 
    The KBC void has been reported to have a characteristic density contrast 
    of $\delta \sim -0.3$ and a scale of $\sim 300\,h_{70}^{-1}\,\mathrm{Mpc}$, 
    corresponding to $z \lesssim 0.07$. 
    The density contrast of the innermost shell obtained in this study 
    is broadly consistent with the value reported for the KBC void \citep{keenan2013}. 
    On the other hand, the innermost shell extends to $z<0.15$, which corresponds 
    to a somewhat wider redshift range than that reported for the KBC void. 
    This difference arises from the simplified top-hat parameterization adopted in this work, 
    in which the boundaries of each shell are determined by the redshift bins of the BAO data. 
    Therefore, the radial extent of this shell does not represent the physical size 
    of the void itself, but rather provides a rough estimate of the density contrast in this region.
    Outside this region, an over-dense region (\bigcnum{2}) is identified, followed by additional under-dense regions (\bigcnum{3} and \bigcnum{4}).
    The regions \bigcnum{5}--\bigcnum{8} are consistent with the background density contrast, $\langle \delta \rangle(z)=0$, at the $1\sigma$ level.

    Such a sequence structure of over- and under-dense regions is also reported by
    wide-field galaxy surveys.
    Ref.~\citep{shimakawa2021}
    analyzed about eight million galaxies with $i<23\,\mathrm{mag}$ from
    the Subaru Strategic Program with the Hyper-Suprime Cam (HSC-SSP),
    based on deep five-band optical photometry ($g,r,i,z,y$) covering
    $\sim360\,\mathrm{deg}^2$ in the redshift range $0.3<z<1$.
    Their analysis revealed several extended under-dense regions with
    density contrasts of $\left<\delta\right> \sim -0.3$ in the range $z=0.3$--$0.6$,
    together with faintly over-dense structures extending to $z\sim1$.
    
    Fig.~\ref{fig:fourfigs}(b) shows the expansion history of the universe over the redshift range $0 \lesssim z \lesssim 2.33$.
    The black solid curve represents the theoretical prediction for $\dot{a}_\mathcal{D}(z) = H_\mathcal{D}(z)/(1+z)$,
    obtained by substituting the spatially averaged matter-density contrast shown in Fig.~\ref{fig:fourfigs}(a) into Eq.~(5),
    while the dashed curve shows the prediction of the flat $\Lambda$CDM model inferred from Planck~2018.
    The labels \bigcnum{1}--\bigcnum{8} correspond to the regions indicated in Fig.~\ref{fig:fourfigs}(a).
    The Hubble tension at $z=0$ is primarily attributed to the presence of the KBC-void-like region (\bigcnum{1}).
    Furthermore, the over-dense region contributes to a reduction of the expansion rate at $z\sim0.3$.
    In the redshift range $0.4 \lesssim z \lesssim 0.8$, the under-dense regions also contribute to an enhanced expansion rate.
    At higher redshifts $(z \gtrsim 0.8)$, the cosmic expansion rates become consistent with the Planck~2018 at the $1\sigma$ level.
    \begin{table}
    \centering
    \caption{
    Results of the reconstructed distribution of the spatially averaged matter-density contrast.
    The labels \bigcnum{1}--\bigcnum{8} correspond to the numbered regions shown in Fig.~\ref{fig:fourfigs}(a).
    }
    \begin{tabular}{c c c}
    \hline
    Structure & Redshift range $z$& $\langle \delta \rangle(z)$ \\
    \hline
    \bigcnum{1} & $0.00 \text{--} 0.15$ & $-0.46 \pm 0.06$ \\
    \bigcnum{2} & $0.15\text{--}0.40$ & $\phantom{-}0.11 \pm 0.05$ \\
    \bigcnum{3} & $0.40\text{--}0.61$ & $-0.16 \pm 0.08$ \\
    \bigcnum{4} & $0.61\text{--}0.82$ & $-0.13 \pm 0.06$ \\
    \bigcnum{5} & $0.82\text{--}1.13$ & $\phantom{-}0.01 \pm 0.04$ \\
    \bigcnum{6} & $1.13\text{--}1.40$ & $\phantom{-}0.02 \pm 0.05$ \\
    \bigcnum{7} & $1.40\text{--}1.91$ & $-0.01 \pm 0.13$ \\
    \bigcnum{8} & $> 1.91$    & $\phantom{-}0.01 \pm 0.04$ \\
    \hline
    \end{tabular}
    \label{tab:voids}
    \end{table}

    Based on the reconstructed spatially averaged matter-density contrast distribution shown in Fig.~\ref{fig:fourfigs}(a),
    we investigate the impact of large-scale structure on the magnitude--redshift ($m$--$z$) relation.
    It has been reported that the apparent-magnitude residuals of Type~Ia supernovae
    remain within an RMS scatter of approximately $0.15$~mag
    with respect to the distance modulus based on a single cosmic expansion rate \cite{brout2022}.
    In contrast, in this study, we interpret the universe defined through spatial averaging over finite domains
    as a sequence of continuously connected effective Friedmann universes,
    each characterized by a different cosmic expansion rate.
    In other words, we emphasize that the inhomogeneous model employed here is not intended to reproduce the $m$--$z$ relation observation over the entire redshift range using a single cosmic expansion history.
    
\begin{figure}
    \centering
    \begin{subfigure}{0.5\textwidth}
    \centering
    \includegraphics[width=\linewidth]{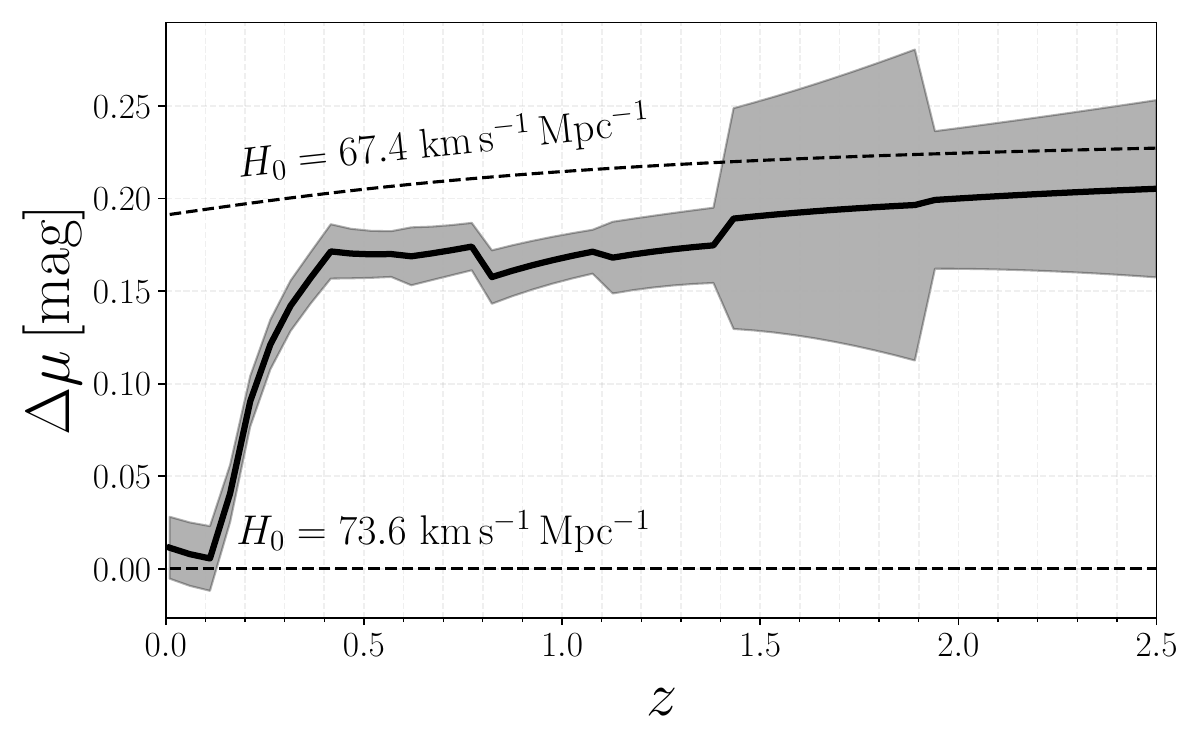}
    \end{subfigure}
    \caption{The apparent magnitude residuals of the inhomogeneous model relative to the Pantheon+ predicted magnitudes.
    The horizontal axis represents the redshift range over which the spatial averaging is performed.
    The gray shaded regions indicate the 68\% (1$\sigma$) credible intervals.
    The dashed line represents the difference in distance modulus $(\Delta\mu)$ for flat $\Lambda$CDM models with $H_0=67.4~\mathrm{km\,s^{-1}\,Mpc^{-1}}$ and $H_0=73.6~\mathrm{km\,s^{-1}\,Mpc^{-1}}$, respectively, relative to the Pantheon+ predicted model.}
    \label{fig:delta_mu}
    \end{figure}

    Ref.~\cite{kenworthy2019} has argued that
    low-density regions do not significantly affect the Type~Ia supernova $m$--$z$ relation.
    However, our model demonstrates that the spatially averaged matter-density contrast
    can induce observable deviations in the apparent magnitudes.
    Fig.~\ref{fig:delta_mu} shows the difference between the distance modulus predicted by the inhomogeneous model
    and that inferred from the Pantheon$+$ Type~Ia supernova data \cite{brout2022}.
    The residual is defined as
    \begin{align}
    \Delta\mu :=5\log
    \frac{
    \displaystyle d_L(\text{Inhomogeneous model})
    }{
    \displaystyle d_L(\mathrm{Flat}\ \Lambda\mathrm{CDM},
    H_0=73.6,\ \Omega_m=0.334)
    } ,
    \end{align}
    where $d_L$ denotes luminosity distance.
    In our model, the $m$--$z$ relation systematically deviates,
    with increasing redshift, from the prediction based on the value
    $H_0 = 73.6~\mathrm{km\,s^{-1}\,Mpc^{-1}}$,
    and gradually approaches the $m$--$z$ relation calibrated by the CMB, corresponding to $H_0 = 67.4~\mathrm{km\,s^{-1}\,Mpc^{-1}}$ \cite{planckcollaboration2020}.
    This indicates that the observed $m$--$z$ relation cannot be fully accounted for by a single cosmic expansion rate.
    Instead, it highlights the importance of defining an effective expansion rate for each spatially averaged domain and interpreting the observables within the spatial-averaging framework.
    A more detailed investigation along this line is left for future work.

\begin{figure}
    \centering
    \begin{subfigure}{0.5\textwidth}
    \centering
    \includegraphics[width=\linewidth]{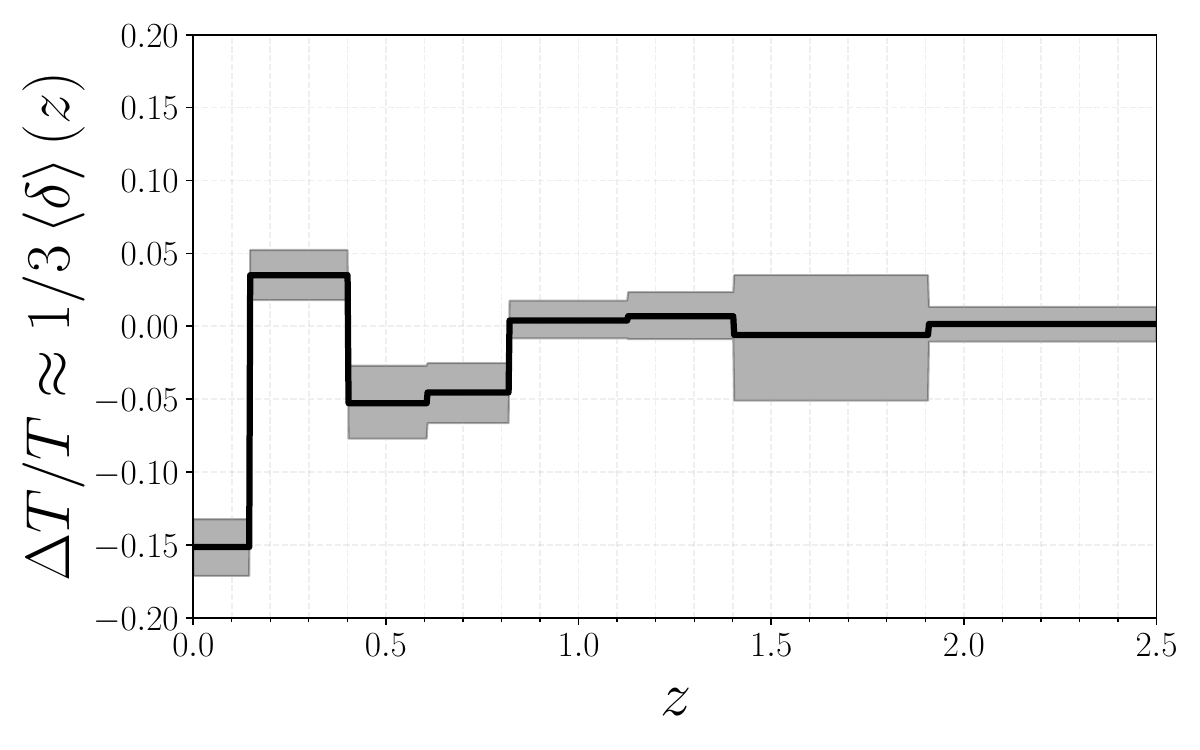}
    \end{subfigure}
    \caption{The order of magnitude of the CMB dipole anisotropy amplitude.
    The horizontal axis represents the redshift range over which the spatial averaging is performed.
    The gray shaded regions indicate the 68\% (1$\sigma$) credible intervals.}
    \label{fig:kSZ}
    \end{figure}

    We consider a rough estimate of the kinematic Sunyaev--Zel'dovich (kSZ) effect induced by the bulk flow associated with large-scale inhomogeneous structures \cite{moss2011}.  
    Following Ref.~\cite{ichiki2016}, the CMB dipole anisotropy amplitude induced by the kSZ effect can be approximated as
    \begin{align}
        \frac{\Delta T}{T} \approx 1 - \frac{a_{\mathcal{D}}(z_\star)}{a(z_\star)}, \label{eq:kSZ}
    \end{align}
    where $a_{\mathcal{D}}(z_\star)$ and $a(z_\star)$ denote the scale
    factors of the spatially averaged domain and the background universe,
    respectively, evaluated at the redshift $z_\star$ at which CMB photons
    enter the inhomogeneous structure.
    Fig.~\ref{fig:kSZ} shows the redshift dependence of $\Delta T/T$.
    If we consider only the innermost top-hat region (region~\bigcnum{1})
    and set the density contrasts of the outer regions
    (regions~\bigcnum{2}--\bigcnum{8}) to zero, the resulting dipole
    amplitude at $z_\star = 0.15$ reaches
    $\Delta T/T \approx -0.15$.
    This value far exceeds the Planck constraint
    $\Delta T/T \lesssim 6.4 \times 10^{-4}$ ($2\sigma$)
    \cite{ade2014},
    indicating that such a simple model based on a single local void is
    strongly disfavored.
    
    On the other hand, if the CMB photons are interpreted as entering the
    line of sight through a much broader spatial region extending to
    $z_\star > 1.91$ (region~\bigcnum{8}), within which the matter density contrast
    is spatially averaged, the predicted dipole amplitude is significantly
    reduced to
    $\Delta T/T \approx 3.3 \times 10^{-3}$.
    This value is substantially smaller than the estimate obtained under
    the single-local-void assumption.
    As the redshift of the inhomogeneous structure through which the CMB photons enter increases,
    the corresponding spatial averaging domain naturally expands.
    As a result, the spatially averaged matter-density contrast
    $\langle \delta \rangle(z)$ is expected to approach zero.
    Consequently, the temperature anisotropy induced by the kSZ effect
    is expected to be suppressed down to the $10^{-4}$ level,
    thereby becoming consistent with the stringent constraints from Planck.

    In addition to the kSZ effect, large-scale inhomogeneities may also
    affect the CMB temperature anisotropies through the integrated
    Sachs--Wolfe (ISW) effect, which arises from the decay of gravitational
    potentials along the photon trajectory.
    For example, Ref.~\citep{kovacs2020} reconstructed the matter-density
    field in the direction of the CMB Cold Spot and identified extended
    supervoid structures around $z\sim0.3$ and $z\sim0.7$.
    Their analysis was performed within the AvERA framework, an
    inhomogeneous cosmological model based on spatial averaging that
    does not invoke a dark-energy component.
    In such models, the dominance of large under-dense regions leads to a
    more rapid decay of gravitational potentials and consequently to
    enhanced ISW signals.
    Although our framework includes a cosmological constant and differs
    from the AvERA scenario in its dynamical assumptions, both approaches
    share the feature that large-scale inhomogeneities are treated within
    a spatially averaged description of the universe.
    In this sense, the presence of comparable large-scale underdensities
    in our reconstructed density profile suggests that similar mechanisms
    may contribute to enhanced ISW imprints.
    
    Finally, if the BAO-inferred expansion history is interpreted as
    arising from the reconstructed matter-density contrast through
    Eq.~(\ref{eq:local_global_hubble}), the density contrasts listed in
    Table~\ref{tab:voids} must also be independently confirmed by galaxy surveys.
    In this interpretation, a $3\sigma$ detection of the deviation of the expansion rate from the
    $\Lambda$CDM prediction inferred from Planck CMB observations is related
    to the spatially averaged density contrast as
    $\Delta H/H = (1/3)f\langle\delta\rangle$, where
    $\Delta H \equiv H_\mathcal{D}-H$.
    This relation implies that the central density contrasts listed in
    Table~\ref{tab:voids} should be measured with a relative precision of approximately
    $30\%$ in order to test the predicted deviation in the expansion rate.
    For the characteristic redshifts corresponding to the transitions in
    Fig.~1(b), this corresponds to uncertainties in the density contrast,
    denoted by $\sigma_{\langle\delta\rangle}$, of approximately
    $\sigma_{\langle\delta\rangle} \sim 0.03$ at $z\simeq0.3$ (region~\bigcnum{2}),
    $\sigma_{\langle\delta\rangle} \sim 0.05$ at $z\simeq0.5$ (region~\bigcnum{3}), and
    $\sigma_{\langle\delta\rangle} \sim 0.04$ at $z\simeq0.7$ (region~\bigcnum{4}).
    
    Future large-volume spectroscopic surveys, such as the DESI five-year dataset and the Prime Focus Spectrograph (PFS) on Subaru \cite{takada2014}, will be crucial for testing whether the reconstructed density profile is consistent with the large-scale matter distribution inferred from
    independent observations.

\section{Conclusion}\label{sec:conclution}
The Hubble tension and the DESI DR2 BAO measurements cast serious doubt on the current understanding of standard cosmology. 
In this study, we demonstrated that large-scale matter inhomogeneities, modeled by the eight top-hat structures, can reconcile the Hubble tension and the expansion history inferred from DESI DR2, without invoking time-varying dark energy. 
We also briefly discussed that a large-scale inhomogeneous universe model based on the spatial-averaging formalism may provide a scenario that can consistently account for the Hubble parameter, the $m$--$z$ relation, as well as the kSZ and ISW effects.
We emphasize that the present study does not attempt to model the detailed impact of inhomogeneities on various cosmological observables. Rather, our results should be regarded as a proof-of-concept demonstration highlighting the importance of reconsidering large-scale matter inhomogeneities. A more quantitative investigation will require further studies using concrete models of spherically symmetric inhomogeneous universes, such as the $\Lambda$LTB model \cite{lemaitre1997,tolman1997,bondi1947}.

\begin{acknowledgments}
TF thanks T.~Murokoshi and M.~Hattori for helpful discussions, which significantly contributed to the preparation of the figures in this paper.
We also thank K.~Ichiki and C.-M.~Yoo for helpful discussions.
RK and MT were supported by the “THERS Make New Standards Program for the Next Generation Researchers” at Nagoya University (JST SPRING, Grant Number JPMJSP2125).
\end{acknowledgments}

\appendix
\section{LTB Solution as a Model for the Spatial Averaging Formalism}\label{sec:LTB}
    In this appendix, we show that the Lemaître--Tolman--Bondi ($\Lambda$LTB) solution,
    which describes a spherically symmetric and radially inhomogeneous spacetime
    \cite{lemaitre1997,tolman1997,bondi1947}, can serve as an effective description of the spatially averaged universe with large-scale inhomogeneous structures.
    By integrating the LTB solution over a finite radial domain $\mathcal{D}$,
    we demonstrate that the matter-density and curvature terms appearing in the Friedmann equation
    can be expressed as volume-averaged quantities over the domain $\mathcal{D}$.
    Although the LTB model predicts an anisotropic expansion rate with different radial and transverse Hubble parameters, these directional dependencies are not present in the spatial averaging. 
    \subsection{Basic equations}
    The metric of the LTB solution is given by
      \begin{align}
    ds^2 = -c^2 dt^2 + \frac{\left(\partial_r R\right)^2}{1 - Kr^2} dr^2 + R^2 \left(d\theta^2+\sin^2\theta d\phi^2\right),\label{eq:LTB_metric}
      \end{align}
    where \(R(t,r)/r\) serves as a scale factor in the $\Lambda$LTB spacetime.
    \(K=K(r)\) is an arbitrary function of the radial coordinate $r$, characterizing spatial curvature in the hypersurfaces in constant-time $\Sigma_t$.
    From the energy-momentum conservation law, \(\nabla_\mu T^{\mu\nu} = 0\), we obtain
      \begin{align}
    \dot{\varrho} + \left( \frac{\partial_r\dot{R}}{\partial_rR} + 2 \frac{\dot{R}}{R} \right)\varrho = 0,
      \label{eq:energy_conservation}
      \end{align}
  with the solution
      \begin{align}
        \varrho=\frac{\rho_{m0}\left(1+\delta(t_0)\right)r^2}{R^2\partial_r R}.\label{eq:rho_solution}
      \end{align}
    We define the Hubble parameters motivated by Eq.~\eqref{eq:energy_conservation} as
    \begin{align}
        H_{\parallel} &:= \frac{\partial_r \dot{R}}{\partial_r R}, 
        \quad 
        H_{\perp} := \frac{\dot{R}}{R},
        \label{eq:hubble_def}
    \end{align}
    where \(H_{\parallel}\) and \(H_{\perp}\) refer to the radial and transverse expansion rates, respectively.
    In the LTB metric, \(H_{\parallel}\) and \(H_{\perp}\) correspond to the radial and transverse expansion rates inferred from the radial BAO \((D_\mathrm{H})\) and transverse BAO \((D_\mathrm{M})\) distances, respectively.
    These two Hubble parameters slightly take different values \cite{camarena2025}.
    However, as will be shown in Eq.~\eqref{eq:Hr_Homega}, within the linear-order spatial-averaging framework, these rates actually become identical.

  Expanding Eq.~\eqref{eq:energy_conservation} to linear order, we can rewrite it as
      \begin{align}
  H_\parallel=3H\left(1-\frac{1}{3}f\delta\right)-2H_\perp.\label{eq:R_Omega}
      \end{align}
  From the Einstein equations, we obtain the local Friedmann equation before spatial averaging in the LTB model, which is given by
      \begin{align}
  \frac{\partial_r\left(R\left(\dot{R}^2+c^2 K\right)\right)}{3R^2\partial_r R} =\frac{8\pi G}{3c^2}\varrho+\frac{c^2 \Lambda}{3}\label{eq:local_Einstein_Eq}
      \end{align}

\subsection{Spatially averaged Einstein equations}
    By integrating Eq.~\eqref{eq:local_Einstein_Eq} over the domain with radius \( r_{\mathcal{D}} \), we obtain the the Friedmann equation over the  domain \( \mathcal{D}\):
    \begin{align}
    H^{2}_{\mathcal{D}\perp}
    := \left( \frac{\dot{R}_{\mathcal{D}}}{R_{\mathcal{D}}} \right)^{2}
    = \frac{8\pi G}{3c^{2}} \frac{M}{R_{\mathcal{D}}^{3}}
    -\frac{c^{2} K\, r_{\mathcal{D}}^{2}}{R_{\mathcal{D}}^{2}}
    + \frac{c^{2}\Lambda}{3}.
    \label{eq:Friedmann_LTB}
    \end{align}
    Here, \( M \) is an arbitrary function related to the gravitational mass given by
    \begin{align}
      M&=\int_{0}^{r_{\mathcal{D}}}3\varrho R^2\partial_r R dr\\
      &=\int_{0}^{r_{\mathcal{D}}}3 \rho_{m0} \left(1+\delta(t_{0},r)\right) r^{2} \, dr .
      \label{eq:total_mass2}
    \end{align}
    Note that we distinguish \( R \) that appears in the Einstein equation before the integration procedure 
    (Eq.~\eqref{eq:local_Einstein_Eq})
    from \( R_{\mathcal D} \) that appears in the spatially averaged Einstein equation
    (Eq.~\eqref{eq:Friedmann_LTB}),
    where
    \( R_{\mathcal D} := R(t, r_{\mathcal D}) \).
    We then consider the scale factor depending on the domain \( \mathcal D \) as
    \( a_{\mathcal D} := R_{\mathcal D}/r_{\mathcal D}, \)
    with normalization \( a_{\mathcal D}(t_0) = 1 \).
    
    Here, we define the spatial average of a scalar quantity \(Q\) over a domain \(\mathcal{D}\) as
    \begin{align}
        \langle Q \rangle
        :=\frac{
            \displaystyle \int_{0}^{r_\mathcal{D}}
            Q\, r^2 dr
        }{
            \displaystyle \int_{0}^{r_\mathcal{D}} r^2 dr
        } .
    \end{align}
    Thus, the spatially averaged matter density is given by
    \begin{align}
        \langle \varrho \rangle
        =\frac{M}{r_\mathcal{D}^3}
        =\frac{
            \displaystyle \int_{0}^{r_\mathcal{D}}
            \rho_{m0}(1+\delta(t_0,r))\, r^2 dr
        }{
            \displaystyle \int_{0}^{r_\mathcal{D}} r^2 dr
        } .\label{eq:averaged_mass}
    \end{align}

    The spatial curvature can be decomposed into a background component and a perturbative contribution arising from the inhomogeneity of the radial density contrast:
    \begin{align}
        K &= K_b + \delta K_\mathcal{D} .
        \label{eq:total_mass1}
    \end{align}
    In what follows, we assume a spatially flat background universe, $K_b = 0$. 
    The curvature perturbation at linear order is obtained as (see subsection~\ref{sec:deltaK} for the detailed derivation)
\begin{align}
    \delta K_\mathcal{D}= \frac{2}{3} \frac{1}{c^2 D_+(t)}
    \frac{
        \displaystyle \int_{0}^{r_\mathcal{D}} Q_+\, r^2\, dr
    }{
        \displaystyle \int_{0}^{r_\mathcal{D}} r^2\, dr
    } .
    \label{eq:carvature}
\end{align}
    As shown in Eqs.~\eqref{eq:averaged_mass} and \eqref{eq:carvature}, the matter density and curvature terms appearing in the Friedmann equation can be expressed as volume-averaged quantities over the domain $\mathcal{D}$.
    Therefore, we obtain the linear-order spatially averaged Friedmann equation as
    \begin{align}
        H_{\mathcal{D}\perp}^2
        = \frac{8\pi G}{3c^2} \frac{\left<\varrho\right>}{a_\mathcal{D}^3}
        - \frac{c^2\, \delta K_\mathcal{D}}{a_\mathcal{D}^2}
        + \frac{c^2 \Lambda}{3}.
        \label{eq:after_average_Fried}
    \end{align}
    This result corresponds to the form shown in Eq.~\eqref{eq:average_Friedman}.
    Similarly, by applying the volume-averaging procedure to the energy conservation law (Eq.~\eqref{eq:R_Omega}),
    we obtain a relation between the radial and transverse Hubble parameters:
    \begin{align}
        H_{\mathcal{D}\parallel}
        = 3H\left(1 - \frac{1}{3} f \langle \delta \rangle \right)
        - 2H_{\mathcal{D}\perp}.
        \label{eq:radi_trans_HD}
    \end{align}
    Following Eqs.~\eqref{eq:average_Friedman} and \eqref{eq:local_global_hubble}, we further find that, at linear order,
    \begin{align}
        H_{\mathcal{D}\parallel}
        = H_{\mathcal{D}\perp}.
        \label{eq:Hr_Homega}
    \end{align}
    As noted in Eq.~\eqref{eq:hubble_def}, the LTB metric generally allows the radial and transverse Hubble parameters
    to take different values.
    However, Eq.~\eqref{eq:Hr_Homega} shows that within the linear-order volume-averaging framework,
    these two expansion rates coincide.
    
\subsection{Derivation of the effective curvature $\delta K_\mathcal{D}$}\label{sec:deltaK}
In this subsection, we derive the effective curvature $\delta K_\mathcal{D}$ in Eq.~\eqref{eq:carvature} using the $3+1$ formalism and linear perturbation theory for matter density fluctuations.
We first summarize the $3+1$ formalism, which describes the evolution of spacetime based on the dynamics of spatial hypersurfaces $\Sigma_t$.

We focus on inhomogeneities arising from linear scalar perturbations on a flat, dust-dominated background.
Assuming matter as a perfect fluid without pressure (dust), the metric is given as
\begin{align}
    ds^2 &= -c^2 dt^2 + \gamma_{ij} dx^i dx^j, \\
    \gamma_{ij} &= a^2 \left( \delta_{ij} + 2 \partial_i \partial_j E + 2 F \delta_{ij} \right), \label{eq:metric_linear}
\end{align}
where $E$ and $F$ denote scalar perturbations.
$\gamma_{\mu\nu} = g_{\mu\nu} + n_\mu n_\nu$ is the projection tensor and $n_\mu = (-1,0,0,0)$ is the unit normal to the hypersurface $\Sigma_t$.
Defining the extrinsic curvature as $K_{ij} := \frac{1}{2c} \dot{\gamma}_{ij}$, the Einstein equations can be decomposed into the Hamiltonian constraint, the momentum constraint, and the evolution equation (trace part):
\begin{align}
\left(K^\ell_\ell\right)^2 - K^i_j K^j_i + {}^{(3)}\mathrm{Ric} &= \frac{16\pi G}{c^4}T_{00} + 2 \Lambda, \label{eq:Einstein00}\\
K^j_{i|j} - K^j_{j|i} &= \frac{8\pi G}{c^4}T_{0i}, \label{eq:Einstein0i}\\
\frac{1}{c}\dot{K}^i_j + K^\ell_\ell K^i_j + {}^{(3)}R^i_j 
      &= \frac{8\pi G}{c^4} \left(T^i_j - \frac{1}{2} T \delta^i_j\right) + \Lambda \delta^i_j,\label{eq:Einsteinij}
\end{align}
where $^{(3)}\mathrm{Ric}$ is the three-dimensional Ricci scalar and $_|$ denotes the three-dimensional covariant derivative.
From Eqs.~(\ref{eq:Einstein00}), (\ref{eq:Einstein0i}), (\ref{eq:Einsteinij}) and the energy conservation law $\nabla_\mu T^{0\mu}=0$, the first-order perturbation equations are obtained:
\begin{align}
    \frac{H}{c^2}\left(\Delta\dot{E}+3\dot{F}\right)-a^{-2}\Delta F &= \frac{4\pi G}{c^4}\rho_b\,\delta, \\
    \dot{F} &= 0, \\
    \frac{1}{c^2}\left(\Delta\ddot{E}+3H\Delta\dot{E}\right)-a^{-2}\Delta F &= 0, \\
    \dot{\delta} + \delta\,\dot{E} + 3\dot{F} &= 0.
\end{align}
Combining these equations, one obtains a second-order differential equation for the density contrast:
\begin{align}
\ddot{\delta} + 2\frac{\dot{a}}{a}\dot{\delta} - \frac{4\pi G}{c^2}\rho_b\,\delta = 0.
\end{align}
Its growing-mode solution is
\begin{align}
  \delta(t,\bm{x}) = D(t)\,Q_+(\bm{x}), \label{eq:delta_eq}
\end{align}
where $D(t) = D_+(t)/D_+(t_0)$.
This leads to the relation
\begin{align}
  \Delta F
  &= -\frac{a^{2}}{c^2}\left(H\dot{\delta}+\frac{4\pi G}{c^2}\rho_b\,\delta\right)\\
  &= -\frac{1}{c^2}\frac{Q_+}{D_+(t_0)}.\label{eq:solution_F}
\end{align}

Introducing first-order perturbations to the scale factor and the three-dimensional spatial metric yields
\begin{align}
    a_\mathcal{D} &= a(1+\epsilon), \label{eq:perturb_aD} \\
    \gamma_{ij}
    &= a^2 \left[
        (1+2\epsilon)\delta_{ij}
        + \left( 2r\frac{d\epsilon}{dr} + \delta K_\mathcal{D} \right)
        \frac{x_i x_j}{r^2}
    \right].
    \label{eq:perturb_LTB}
\end{align}
From Eqs.~\eqref{eq:solution_F}, \eqref{eq:perturb_aD}, and \eqref{eq:perturb_LTB}, the curvature perturbation at linear order is
\begin{align}
    \delta K_\mathcal{D}
    &= -\frac{2}{r}\frac{dF}{dr}\bigg|_{r_\mathcal{D}}
    \notag \\
    &= \frac{2}{3} \frac{1}{c^2 D_+(t)}
    \frac{
        \displaystyle \int_{0}^{r_\mathcal{D}} Q_+\, r^2\, dr
    }{
        \displaystyle \int_{0}^{r_\mathcal{D}} r^2\, dr
    } .\label{eq:derive_Keff}
\end{align}
This result corresponds to the form shown in Eq.~\eqref{eq:Keff}.

\bibliography{Cosmology}
\end{document}